\documentclass[aps,prl,twocolumn,groupedaddress,10pt]{revtex4-1}
\usepackage{graphicx}
\usepackage[latin1]{inputenc}
\usepackage{grffile}

\usepackage{amsfonts}
\usepackage{amsmath}
\usepackage{amssymb}
\usepackage{color}
\usepackage[normalem]{ulem}
\usepackage{todonotes}
\usepackage[bookmarks, 
            breaklinks, 
            pdfstartview=FitH, % fit width when the document is opened
            pdftitle={BECBeamsplitter},
            pdfauthor={T.~Berrada et al.},
            pdfkeywords={}
            ]{hyperref}

\begin{document}

%\title{Recombination beam-splitters for trapped Bose-Einstein condensates}
\title{Matter-wave recombiners for trapped Bose-Einstein condensates}
\author{T.~Berrada}
\author{S.~van~Frank}
\author{R.~Bücker}
\email[]{current affiliation: Max Planck Institute for the Structure and Dynamics of Matter, CFEL, Luruper Chaussee 149, 22761 Hamburg, Germany}
\author{T.~Schumm}
\author{J.-F.~Schaff}
%\email[]{jschaff@ati.ac.at}
\author{J.~Schmiedmayer}
\email[]{schmiedmayer@atomchip.org}
\affiliation{Vienna Center for Quantum Science and Technology, Atominstitut, TU Wien, Stadionallee 2, 1020 Vienna, Austria}
\author{B.~Julía-Díaz}
\author{A.~Polls}
\affiliation{Departament d'Estructura i Constituents de la Matèria and Institute de Ciències del Cosmos, Facultat de Física, Universitat de Barcelona, E-02028, Barcelona, Spain}

\date{\today}

\begin{abstract}
Interferometry with trapped atomic Bose-Einstein condensates (BECs) requires  the development of techniques to recombine the two paths of the interferometer and map the accumulated phase difference to a measurable atom number difference. We have implemented and compared two recombining procedures in a double-well based BEC interferometer. The first procedure utilizes the bosonic Josephson effect and controlled tunneling of atoms through the potential barrier, similar to laser light in an optical fibre coupler. The second one relies on the interference of the reflected and transmitted parts of the BEC wavefunction when impinging on the potential barrier, analogous to light impinging on a half-silvered mirror. Both schemes were implemented successfully, yielding an interferometric contrast of $\sim 20\%$ and 42\% respectively. Building efficient matter wave recombiners represents an important step towards the coherent manipulation of external quantum superposition states of BECs.
\end{abstract}

\maketitle

The most striking application of the wave character of matter is the construction of matter-wave interferometers~\cite{Cronin2009}. Matter-wave interferometry relies on 1) the splitting of the atomic wavefunction between two internal and/or external states with a well-defined phase difference, 2) the possibility 
to implement an additional phase shift during the time evolution, and 3) the recombination of the two wave packets in order to transform their relative phase difference into a measurable signal. The coherent manipulation of atoms in particular has required the development of an atom optics toolbox of beam-splitters, phase shifters, recombiners, etc.

Various interferometric schemes have been devised for BECs either using radio-frequency (rf) or microwave fields to perform a Ramsey 
sequence~\cite{Hall1998,Minardi2001,Chevy2001,Gross2010,Altin2011,Egorov2011,Petrovic2013,Machluf2013}, 
or laser fields to drive Raman~\cite{Doring2010} or Bragg~\cite{Torii2000,Simsarian2000,Denschlag2000,Gupta2002,Hellweg2003,Wang2005,Horikoshi2006,Garcia2006,Debs2011,McDonald2013,Muntinga2013,Kuhn2014} transitions. Most of these schemes resort to free-falling clouds, which inherently limits the interrogation time to a few 100 ms (with the notable exception of experiments conducted in microgravity~\cite{Muntinga2013}). 

Interferometers where atoms are confined in a potential until readout~\cite{Shin2004,Schumm2005,Stoferle2004,Hadzibabic2006,Berrada2013,VanFrank2014} can be used as a $µm$-sized scanning probe, e.g. for high-resolution field sensing~\cite{Bohi2010,Ockeloen2013} or the study of short-ranged interactions. In principle, they also offer the perspective of arbitrarily long interrogation times. However, the effect of interactions can generally not be neglected in trapped-atom interferometers, particularly when working with BECs. On the one hand, interactions cause mean-field shifts and dephasing effects that limit the interrogation time~\cite{Anderson2009,Horikoshi2006,Jo2007,Berrada2013}, but on the other hand they can be used to produce non-classical states and perform measurement with improved sensitivity~\cite{Gross2010,Riedel2010}. Interferometers relying on superpositions of external modes, albeit technically challenging, are particularly relevant for technological applications related to precision measurement of gravitation or inertial forces~\cite{Clauser1988,Peters1999,Gustavson2000}. 

%A fundamental feature of BECs is the role of atom-atom interactions. 
%While they can usually be neglected in freely expanding atomic 
%clouds, interactions dominate the physics of confined BECs, 
%leading to mean-field shifts and dephasing effects, which ultimately 
%limit the coherence time of interferometers~\cite{Anderson2009,Horikoshi2006,Jo2007,Berrada2013}. 
%The impact of interactions is particularly significant in interferometers 
%where the condensates are prepared in a superposition of spatial 
%modes by manipulating the 
%confining potential~\cite{Shin2004,Schumm2005,Stoferle2004,Hadzibabic2006,Berrada2013,VanFrank2014}.

\begin{figure}[t] %\hspace*{-10px}
\centering
\includegraphics[width=\linewidth]{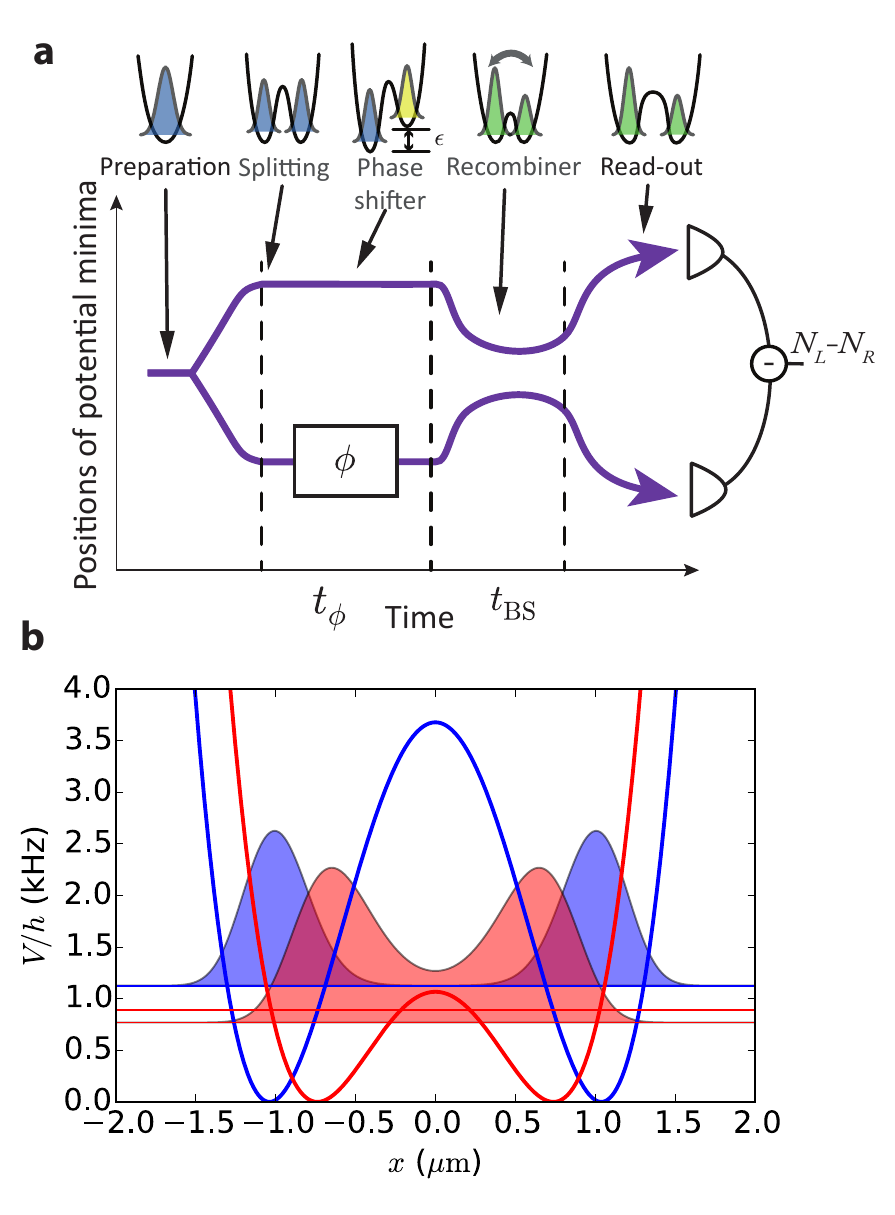}%
\caption{ \textbf{Scheme of the interferometer and double-well potentials.} \textbf{(a)} The condensate is coherently split by transforming a
single trap into a double-well potential; a relative phase between the two arms is imprinted by tilting the double well during a time $t_\phi$ ; the spacing between
the two wells is then reduced to perform the recombination of the wavepackets and map the relative phase onto a
population imbalance. After the recombination time $t_\mathrm{BS}$ , the atom clouds are separated and the atom number in each well is counted. \textbf{(b)} Cut of the double-well potential used immediately prior recombination (thick blue line) and that used for 
the recombiner (thick red line). Thin horizontal lines: chemical potential (including zero-point energy) in the ground and first excited state in each potential (for the weakly coupled potential, the spacing between the levels is smaller than the width of the lines). Shaded surfaces: density profiles of the ground state in each potential. \label{Fig:SchematicsPotentials}}
\end{figure}
	
%\paragraph{\bf Recombiner}

An important class of trapped-atoms interferometers are double-well interferometers. In double-well interferometers~\cite{Shin2004,Schumm2005,Berrada2013}, the splitting into two localized spatial modes is achieved by smoothly transforming a single-well potential into a double-well potential. The inverse operation, namely the recombination of the two modes in order to transform the phase difference between the two paths into a measurable signal, turns out to be more strenuous. While several schemes for the splitting and merging of clouds of thermal atoms have been proposed~\cite{Hansel2001a,Hansel2001b,Andersson2002}, trapped BEC interferometers usually rely on the time-of-flight (tof) recombination method, already used in Refs~\cite{Andrews1997,Schumm2005b}, which implements a matter-wave equivalent of the double-slit experiment. With this technique, the phase information is deduced from the position of the emerging interference fringes in the spatial density distribution of the overlapping atomic clouds. Alternative techniques have been suggested to infer the phase between trapped BECs, e.g. through the phase-dependent excitations produced at the merging of two condensates~\cite{Jo2007b,Negretti2004} or by monitoring the time-evolution of their momentum distribution~\cite{VanFrank2014}. However, no direct method has been demonstrated so far to map the phase of a superposition of external states onto a measurable atom number difference, in contrast to interferometers relying on internal-state labeling, where such techniques are readily available. Developing such a method would make phase-estimation in BEC interferometers with external-state superpositions much more easy since counting atom numbers in two spatial modes is more robust and less sensitive to noise than fitting spatial interference fringes, as it does not requires high spatial resolution imaging nor the measurement of higher-order spatial correlation functions~\cite{Chwedenczuk2010}.

Here, we present two novel methods for the phase-sensitive recombination of the two halves of a BEC, trapped in a double-well potential, implemented on our atom-chip based interferometer~\cite{Berrada2013}. In an optical Mach-Zehnder interferometer (MZI), this operation is realized by means of a recombination beam splitter. Its function is to transform the phase difference between the two paths of the interferometer into a measurable intensity difference between the two output ports. Similarly, we perform this operation by carefully manipulating the confining potential to recombine the two halves of the trapped BEC in such a way as to translate their relative phase directly into an atom number difference between the two wells.

The first method (``Josephson recombiner") relies on the atomic Josephson effect in the double-well potential~\cite{Albiez2005,Pezze2005,Pezze2006,Grond2011}. A phase difference between the two halves of the BEC induces an oscillating tunneling current (Josephson oscillations) through the potential barrier. To control this current, the trap is smoothly deformed so that the wavefunction essentially remains in a superposition of the two lowest-lying modes of the double-well potential at all times. This technique implements a matter-wave analogue to an optical fibre coupler. 

The second method (``non-adiabatic recombiner") is based on the interference of the parts of the BEC wave functions which are reflected on and transmitted through the potential barrier, similarly to a half-silvered mirror in optics. In this case, the fast transformation of the potential with respect to the timescale of the motion in the trap implies that many modes are excited.

In contrast to the tof recombination technique usually employed in double-well interferometers, here the phase is inferred from the measured atom numbers in each well, allowing to draw benefit from the precise atom counting methods already available~\cite{Esteve2008,Bucker2009,Lucke2011}.

Another fundamental difference with the tof recombiner is that atoms remain trapped at all times, making it possible to resort to on-chip detection techniques~\cite{Heine2010,Volz2011}. This also implies that atomic interactions play a key role, and in particular that they can be taken advantage of to engineer the atom number fluctuations of the output state. This should enable interferometric phase inference with a better precision than with the tof recombination~\cite{Chwedenczuk2012}.

\subsection*{\bf Setup and methods}

The basics of the BEC Mach-Zehnder interferometer have been described in our previous publication Ref.~\cite{Berrada2013}. Here we 
summarize the procedure used to prepare the BEC in a superposition of the left and right spatial modes in a double well with a well-controlled relative phase (see Fig.~\ref{Fig:SchematicsPotentials}a). We prepare a BEC with $N \approx 1200$ $^{87}\mathrm{Rb}$ atoms at a temperature $T\approx25$\,nK in an elongated magnetic trap, created using an atom chip~\cite{Trinker2008b}. We use the technique of rf dressing~\cite{Lesanovsky2006,Hofferberth2006a} to dynamically transform the single-well potential into an elongated double well with tunable spacing, barrier height and tilt. By increasing linearly the rf amplitude, we smoothly split the BEC transversely into a symmetric double well~\cite{Schumm2005} with well spacing $d=2\,\mathrm{\mu m}$, barrier height $V_0 = h \times 3.7$\,kHz and tunnel coupling energy $J = h \times 0.1$\,Hz (see Supplemental Material (SM), $h$ denotes Planck's constant). The splitting produces a coherent superposition with a reproducible initial phase difference $\langle \phi_0 \rangle = 0$ (standard deviation of the initial phase difference $\Delta \phi_0 = 0.16\,$~rad) and no population imbalance on average $\langle z  \equiv  (N_L-N_R)/N \rangle= 0$ (the brackets denote ensemble averaging).  To prepare a state with a finite relative phase, we slightly tilt the double well off the horizontal plane, inducing a deterministic phase shift due to the potential energy difference between the two wells
\begin{align}
\phi(t_\phi) = \varphi_\text{0} + \epsilon \, t_\phi/\hbar .		
\end{align}
with $\epsilon/h=350$\,Hz~\cite{Baumgartner2010}. The phase of the superposition can be adjusted by tuning the phase 
accumulation time $t_\phi$. $\varphi_0$ is a phase offset picked up while the double well is tilted and levelled back, and hence independent of $t_\phi$.
%The tilt is performed in 3\,ms, and 
%eventually reversed in another 3\,ms. During this time, the phase 
%also evolves by a fixed amount $\phi_0$ independent from $t_\phi$. 

The linear evolution of the mean phase is accompanied by a broadening of the phase distribution caused by atomic interactions which is currently the main limitation to the interrogation time of the interferometer~\cite{Lewenstein1996,Berrada2013}. Because of the difference of chemical potential between different states with well-defined atom number difference, the initial relative number uncertainty after splitting translates into a growing relative phase uncertainty, with the variance of the relative phase growing as~\cite{Castin1997,Javanainen1997,Leggett1998}
\begin{equation}
\Delta \phi^2 (t_\phi) = \Delta \phi_0^2 + R^2 (t_\phi - t_\text{i})^2 \,.
\label{Eq:PhaseDiff}
\end{equation}
The dephasing rate $R=51\,\mathrm{mrad\cdot ms^{-1}}$ is proportional to the interaction energy and to the initial uncertainty on the population imbalance after splitting, and $t_i=-6$\,ms accounts for the two times 3\,ms used to incline and level the double 
well. 

To characterize the state of the BEC, we can interrupt the sequence at any time and measure either the relative phase $\phi$ or the population imbalance $z$. To infer $\phi$, we switch off the trap, let both halves of the condensate overlap and image them with our tof fluorescence imaging system~\cite{Bucker2009}. We extract the phase from the position of the interference fringes in the density profile of the expanded cloud with an estimated $1\sigma$-uncertainty of $\pm0.08$~rad. To measure $z$, we switch off the trap in such as to apply a kick with opposite momentum to each cloud, and count the atoms in two separate regions of the fluorescence pictures. We estimate the $1\sigma$-uncertainty of the atom number difference measurement to be of the order of $\pm13$~atoms~\cite{Berrada2013}.

Altogether, the splitting and the phase accumulation stages produce a coherent superposition of left and right modes with a reproducible mean phase, and a phase spread which increases in time under the effect of interaction-induced dephasing.

\begin{figure*} %\hspace*{-10px}
\centering
\includegraphics[width=0.7\linewidth]{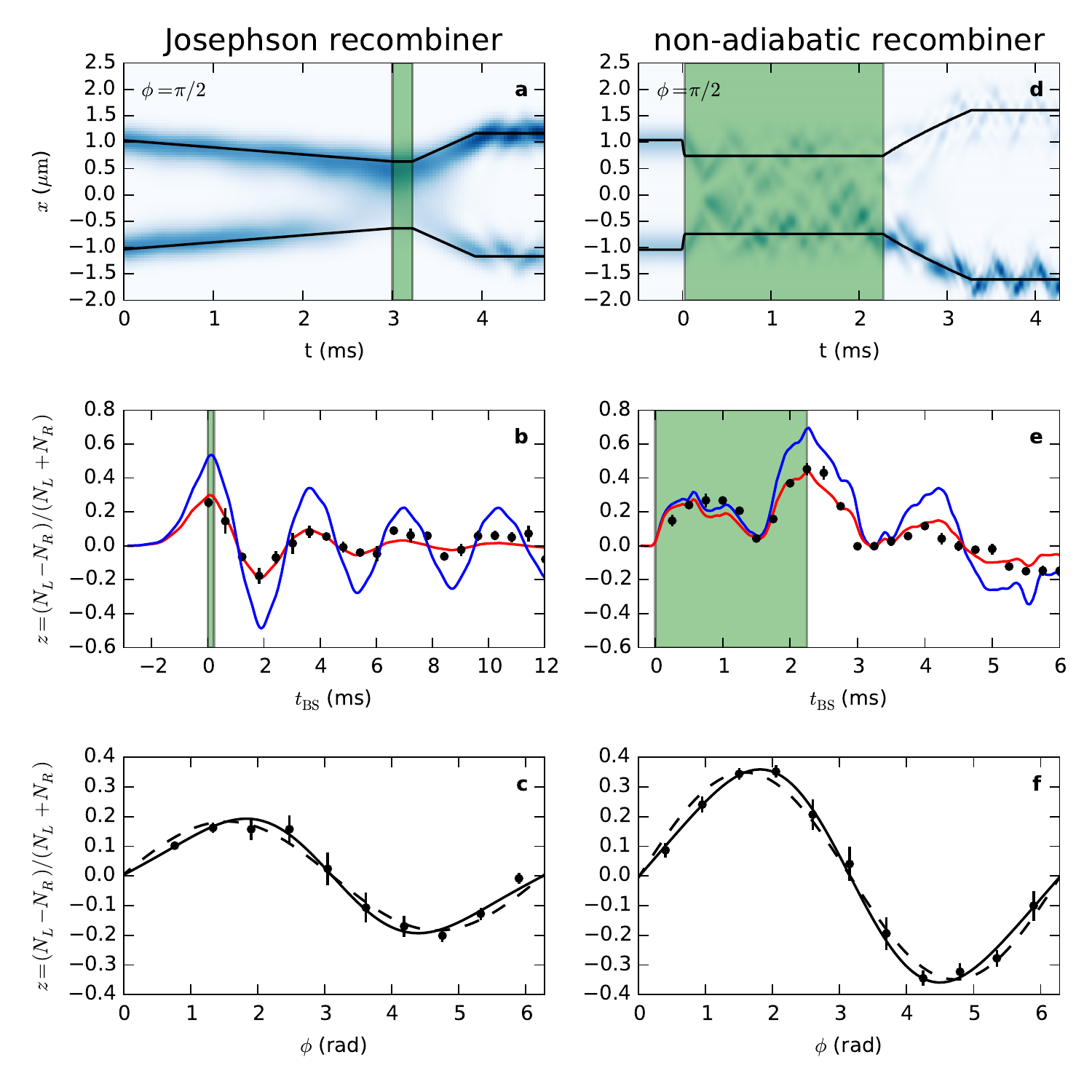}%
\caption{\textbf{The two BEC recombiners.} Left: Josephson recombiner, right: non-adiabatic recombiner. \textbf{a,d} : Simulated evolution of the density profile in the direction of splitting during the beam-splitter operation, for an initial state with $z=0, \phi_i=\pi/2$. Black lines: instantaneous position of the double-well minima, green shaded area: holding time $t_\mathrm{BS}$ in the recombination double well. Times below 0 ms refer to the coupling ramp. \textbf{b,e} : Evolution of the final imbalance as a function of $t_\mathrm{BS}$. Black points: measured imbalance (ensemble average), blue line: result of a 3D Gross-Pitaevskii simulation, red line: same as blue line time-shifted and multiplied by an exponential damping term to fit the data. Decay times:  5.1~ms (Josephson recombiner), 5.4~ms (non-adiabatic recombiner). The shaded area correspond to the experimentally optimized value of $t_\mathrm{BS}$ yielding the maximal output imbalance.  \textbf{c,f} : Final imbalance as a function of the relative phase at input for the optimal $t_\mathrm{BS}$. Dots: experimental data, continuous line: fit with two-harmonic model, dashed line: harmonic part of the two-harmonic fit. The error bars indicate $\pm$ one std. error of the mean.
\label{Fig:BSCalib}}
\end{figure*}

\subsection*{\bf Recombiners}

The last element needed to close the interferometric sequence is 
a phase-sensitive recombiner. We have implemented two methods to perform the phase-dependent recombination of the two halves of the BEC, i.e.\ to transform a symmetric superposition of the two modes with a phase difference $\phi(t_\phi)$ into a state with a population imbalance $z$ depending on $\phi(t_\phi)$. Both rely on the coherent motion of the atoms in the double-well potential. In the following, we describe the details of each protocol.

\paragraph{\bf Josephson recombiner.}

\begin{figure}[h] %\hspace*{-10px}
\centering
\includegraphics[width=\linewidth]{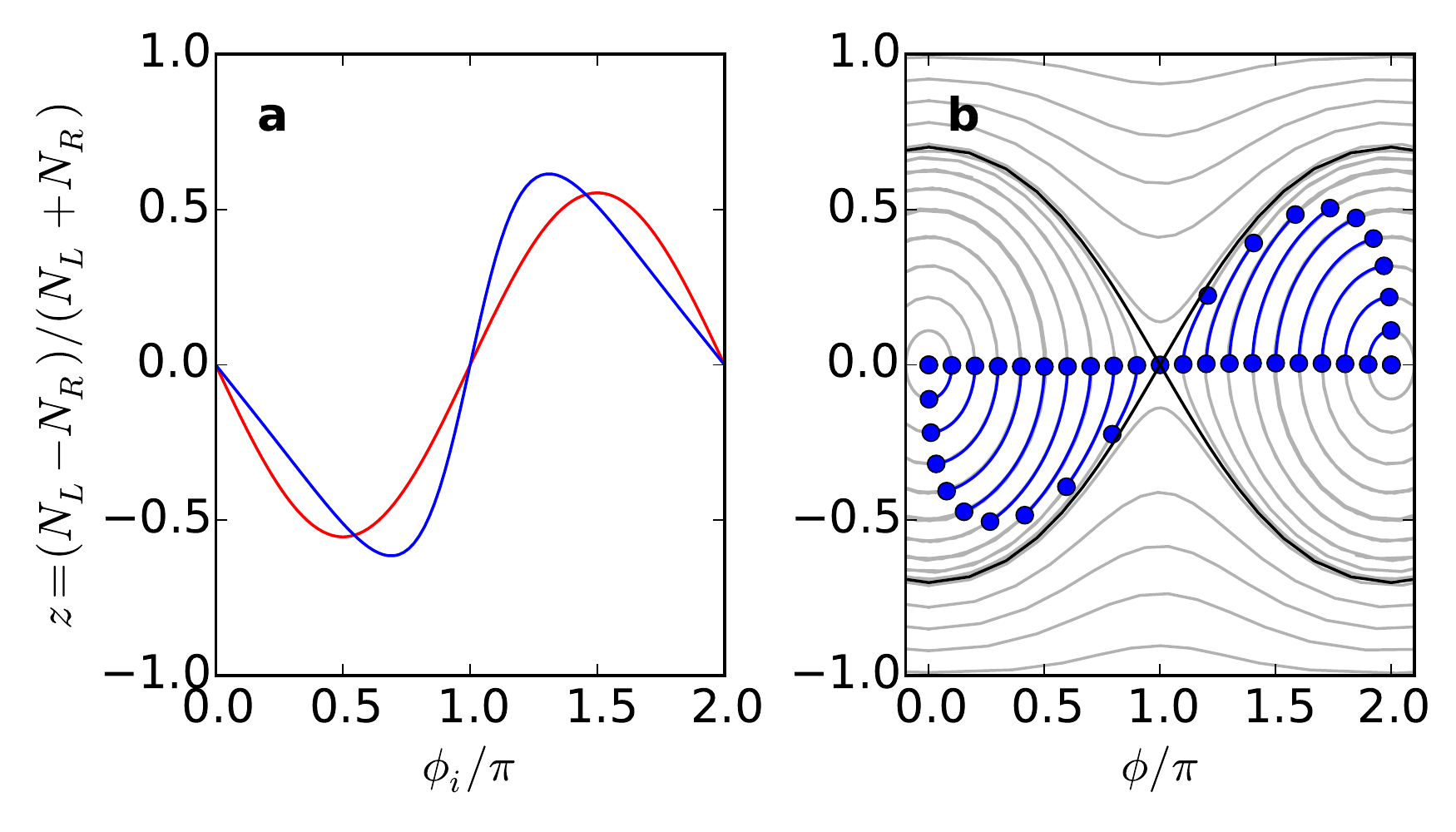}%
\caption{ \textbf{Effect of interactions.} \textbf{a:} Final population 
imbalance $z$ as function of the phase $\phi_i$ of the state at the input 
of the Josephson recombiner, in absence (red) and in presence (blue) of 
interactions (1D GPE simulation of the recombiner sequence along the direction of splitting ). Interactions 
are responsible for the anharmonicity of the blue curve. Note the steep 
slope close to $\phi_i=\pi$. \textbf{b.} Classical phase portrait of 
the BJJ for $\Lambda=10$ (gray lines). The blue points on the $z=0$ axis 
correspond to input states of the recombiner with different initial phases. 
The blue lines represent the trajectory each state travels in phase space 
during a fixed time equal to a quarter of a (small-amplitude) plasma 
oscillation. As the initial state gets closer from 
the hyperbolic fixed point $(\phi=\pi,z=0)$, the oscillations become 
increasingly slow and approach asymptotically the separatrix (black line). 
\label{Fig:JBSSimInt}} 
\end{figure}	
\begin{figure*}[h!] %\hspace*{-10px}
\centering
\includegraphics[width=0.9\linewidth]{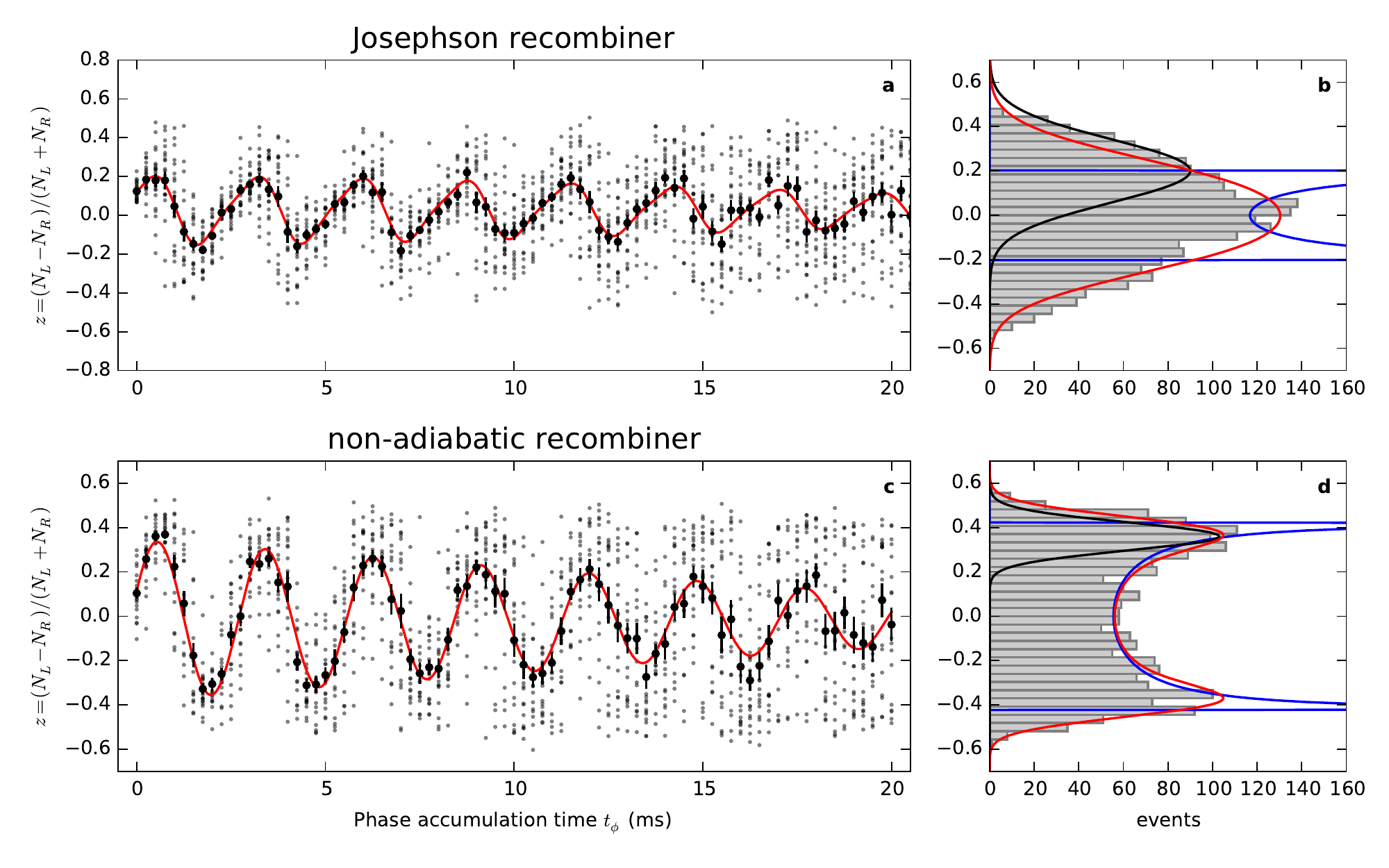}
\caption{\textbf{Mach-Zehnder interferometric fringes.} \textbf{Left:} Normalized population imbalance between the two wells measured as a function of the phase accumulation time prior to recombination with the Josephson (\textbf{a}) or non-adiabatic (\textbf{c}) recombiner. It exhibits interference fringes and a damping due to interaction-induced dephasing. Gray dots: imbalance of individual realizations, black dots: ensemble average, red curve: fit with a model taking into account dephasing. Note that the first oscillation for each recombiner corresponds to the data of Fig.~\ref{Fig:BSCalib} c and f. \textbf{Right:} Distribution of population imbalance over all times obtained by binning all the single-realization imbalances from the Josephson BS (\textbf{b}) and the non-adiabatic BS (\textbf{d}) data, used to extract the contrast of the recombiners. Red line: Fit with the model of~Eq.~\eqref{Eq:BatmanDis} (blue line) with additional Gaussian imblanance noise (black line). \label{Fig:MZIfringes}} 
\end{figure*}

A natural way to translate the input phase difference into a population imbalance is to make use of the atomic Josephson effect in the double-well potential~\cite{Albiez2005,Pezze2005,Pezze2006,Grond2011}. Assuming that the BEC wavefunction can be written as a superposition of two time-independent spatial modes in a symmetric double-well potential, the time-evolution of the population imbalance and relative phase obey the coupled equations
\begin{align}
		&\dot{z} = - \sqrt{1-z^2(\tau)} \sin \phi (\tau) \label{Eq:zdotnodim}\\
		&\dot{\phi} = \Lambda z(\tau) + \frac{z(\tau)}{\sqrt{1-z^2(\tau)}} \cos \phi (\tau). \label{Eq:phidotnodim}
\end{align}
where $\tau = 2Jt/\hbar$ is a dimensionless time rescaled to the single-particle tunneling frequency $J/h$, and $\Lambda = UN/2J$ denotes the ratio of interaction to tunneling energy (see SM for the definition of the parameters)~\cite{Raghavan1999}. 

In absence of interaction ($\Lambda=0$), starting from a state with $z=0$ and a given initial relative phase $\phi_i$ and letting the atoms tunnel for a quarter of an oscillation period produces a state with imbalance
\begin{align}
	z(\phi_i) = \sin \phi_i.
\end{align}
This is equivalent to a $\pi/2$ Rabi pulse in Ramsey interferometry.

The Josephson recombiner is implemented by ramping down the double-well barrier in 3~ms from the split trap to a more coupled trap (see Fig.~\ref{Fig:SchematicsPotentials}b). The duration of the coupling ramp was chosen to avoid exciting a sloshing motion of the BEC. The condensates are then held for an adjustable time $t_\mathrm{BS}$ in the coupled trap, before the barrier is raised again to separate the atoms for counting. The procedure is illustrated in Fig.~\ref{Fig:BSCalib}a.

The parameters of the recombiner (recombination double well and duration of the holding time $t_\mathrm{BS}$) were experimentally optimized to achieve the maximum output imbalance starting with a symmetric input superposition with a phase close to $\pi/2$. For each final double well, we scanned the duration of the holding time $t_\mathrm{BS}$ and monitored the Josephson oscillations (Fig.~\ref{Fig:BSCalib}b). We obtained the highest population imbalance $\left\langle z_m\right\rangle \approx 0.2$ in the potential displayed in Fig.~\ref{Fig:SchematicsPotentials}b (thick red line) for $t_\mathrm{BS}=0.225$~ms. In this double well, the distance between the two potential minima is $1.5\,\mathrm{\mu m}$ and the barrier height is $h\times1$\,kHz. Mean-field simulations of the tunneling dynamics show that most of the tunneling already occurs during the recombination ramp, while the two potential wells are being coupled. From the measured frequency of the Josephson oscillation, we estimate $J/h\approx 40$~Hz and $\Lambda \approx 7$. 

By scanning the phase-accumulation time $t_\phi$ to tune the relative phase at the input of the recombiner, we observe a sine-like dependence of the output imbalance (see Fig.~\ref{Fig:BSCalib},~c). This oscillation exhibits a characteristic anharmonic shape, with a slope steeper around $\phi_i=\pi$ than around $\phi_i=0$. The anharmonicity is caused by atomic interactions and is already captured by the classical two-mode description of the BEC (Eqs.~\eqref{Eq:zdotnodim} and \eqref{Eq:phidotnodim}). Starting from a state with no imbalance and varying the initial relative phase, the period of the oscillations of $\phi$ and $z$ around the stable point $(z=0,\phi=0)$ diverge as one gets closer from the separatrix between the Josephson oscillations and the self-trapped modes (thick black line in Fig.~\ref{Fig:JBSSimInt}b)~\cite{Raghavan1999}. Hence, for a state with an initial phase $\phi_i \approx \pi$, a small variation of $\phi_i$ causes a large variation of the imbalance measured after a fixed duration $t_\mathrm{BS}$ equal to a quarter of a small-amplitude plasma oscillation. Interestingly, this anharmonicity in the vicinity of $\phi_i=\pi$ resembles the non-linearity predicted in Ref~\cite{Negretti2004} at the merging of two condensates, which had been suggested to improve the phase sensitivity.

As result of the parity of the response of the recombiner and the $2\pi$-periodicity of the phase, the output imbalance at a given time can be written as a Fourier series
\begin{align}
	z(\phi_i) = \sum_{n=1}^M a_n \sin \left( n \phi_i \right) \label{Eq:Anharmonic}
\end{align}
without loss of generality. Fitting the data of Figure~\ref{Fig:BSCalib}c with the model of Eq.~\eqref{Eq:Anharmonic}, we find fair agreement already by restricting the series to the two lowest harmonics with $a_1=0.18\pm0.02$ and the anharmonicity $\eta \equiv \left|a_2/a_1\right|=0.26\pm0.13$. As long as the lowest harmonic $abs\left|a_1\right|$ dominates over the higher-order harmonics, the amplitude $a_1$ of the recombiner response is an indication on how sensitive it is to phase shifts. However, a higher phase uncertainty on the input state will tend to reduce this amplitude (as one has to average over different output imbalances) independently from the intrinsic contrast of the recombiner.

To characterize the Josephson recombiner regardless of the input state, we estimate its intrinsic contrast. It corresponds to the highest achievable output imbalance $\left|z(\phi_i)\right|$ when varying the phase $\phi_i$ of the input state. In practice, the phase of the input state is tuned by varying the phase accumulation time $t_\phi$. The corresponding imbalance at the output of the recombiner displays the Mach-Zehnder interference fringes represented in Fig.~\ref{Fig:MZIfringes}a. However, for a given value of $t_\phi$, the phase exhibits some uncertainty, which grows with $t_\phi$ under the effect of interaction-induced dephasing, as modelled in Eq.~\eqref{Eq:PhaseDiff}. When averaging over experimental realizations for each value of $t_\phi$, this phase noise reduces the amplitude of the interference fringes. To estimate the contrast $C$ of the recombiner independently from the uncertainty on the phase of the input state, we resort to the method presented in Ref.~\cite{Berrada2013} : we construct the distribution of population imbalance by binning the measured single-shot imbalances at all times and assume that the phase uniformly samples the interval $\left[0,2\pi\right]$. Neglecting the anharmonicity of the recombiner, we expect the output imbalance $z$ to be distributed following
\begin{align}
	p(z) &= \frac{1}{\pi}\frac{1}{\sqrt{z^2-C^2}} \quad \mathrm{if}~\left|z\right|<C, \nonumber \\
	 & = 0 \quad \mathrm{elsewhere,} \label{Eq:BatmanDis}
\end{align}
and thus to exhibit a typical two-peak structure (blue line in Fig.~\ref{Fig:MZIfringes}b and d). To account for technical atom-number noise of the recombiner, the distribution Eq.~\eqref{Eq:BatmanDis} is convolved with a Gaussian distribution of rms width $\sigma_\mathrm{rec}$. We compare the distribution of imbalances to this model to estimate $C$ and $\sigma_\mathrm{rec}$. Unfortunately, the measured distribution does not display a two-peak structure, which makes it difficult to fit with our model. Imposing $C=0.2$, we find rough agreement with the data for $\sigma_\mathrm{rec} =0.15$. We attribute the failure of the fit model to the relatively low contrast of the Josephson recombiner as well as its anharmonicity. However, even including the anharmonicity in the model suggests that $C \lesssim 0.2$ and $\sigma_\mathrm{rec} \gtrsim 0.15$, both parameters being strongly anti-correlated.

The contrast of the Josephson recombiner is fundamentally limited by the onset of self-trapping~\cite{Raghavan1999}. In a simple two-mode picture, the maximum imbalance achievable when starting with $z=0$ is set by the self-trapping threshold
\begin{align}
\left|z_c\right| = 2\frac{\sqrt{\Lambda-1}}{\Lambda} \label{Eq:MQSTthreshold} 
\end{align}
(see Fig.~\ref{Fig:JBSSimInt}b). For $\Lambda \sim 7$, we expect $\left|z_c\right| \approx 0.7$, which is much larger than the highest achieved imbalance. This suggests that a recombiner based on the Josephson effect should operate in the Rabi regime ($\Lambda < 1$), where tunneling dominates over atomic interactions. However, attaining this limit with our setup seems strenuous, since even for the most strongly coupled double wells, the interaction energy $UN \approx \mu \approx h \times 0.5$\,kHz is larger than or equal to the spacing between the two lowest energy levels. Furthermore, approaching the Rabi regime implies strongly reducing the splitting distance, and hence significantly displacing the condensate wave function. Achieving adiabaticity with respect to the motion of the atoms implies deforming the 
potential over long timescales, of the order of tens of ms, during which 
other effects limit the coherence of the superposition. Faster manipulation of the atoms implies breaking the adiabaticity, and hence invalidates the two-mode description of the recombiner.

Another factor limiting the contrast is the way we separate the wavepackets to measure the imbalance: ideally, we wish to adiabatically map the output state of the recombiner onto a superposition of the left and right modes of two uncoupled wells. In practice however, separating the two halves of the BEC too abruptly is similar to projecting its wavefunction \textit{prior} separation onto the left and right modes \textit{after} separation, which effectively reduces the final measured imbalance, as can be seen in Fig.~\ref{Fig:JBSSimInt}a : even in absence of interaction, the output imbalance, defined as $z=\int_{-\infty}^0 \left|\psi(x) \right|^2\mathrm{d}x - \int^{\infty}_0 \left|\psi(x) \right|^2\mathrm{d}x$ is only of the order of 50\% for an input phase $\phi_i=\pi/2$  (here $x$ denote the direction of splitting in the double well, and $\int_{-\infty}^{\infty} \left|\psi(x) \right|^2\mathrm{d}x=1$).

To estimate what contrast should be theoretically achievable, we simulated the dynamics of the BEC during the whole recombiner sequence by solving the 3D Gross-Pitaeevski equation (GPE). The simulations indicate that contrasts as high as $\sim 55\%$ should be attained. We attribute this discrepancy with the best measured contrast of $\sim 20\%$ to effects beyond the mean-field picture. Possibly, the limited contrast is related to the strong damping of the tunneling oscillations that we systematically observe in our double wells (see Fig.~\ref{Fig:BSCalib}b), and which will be the object of further studies. To quantify the effect of this damping, we fit the the data of Fig.~\ref{Fig:BSCalib}b with the result of a 3D GPE simulation of the dynamics of the BEC with an adjustable time shift $\Delta t$ (to account for a small experimental delay) and an exponential damping prefactor $e^{-t/\tau}$, yielding $\Delta t=0.15$~ms and $\tau=5.1$~ms. Note that the  3D GPE simulation also exhibits a damping (blue line in Fig.~\ref{Fig:BSCalib}b), but on a much longer timescale than observed experimentally. On an even longer timescale, the simulation predicts revivals of the amplitude which we never observed experimentally. The short timescale (with respect to the Josephson oscillation period $T_J = 3.8$\,ms) of this unknown additional damping mechanism emphasizes the need for a fast recombination procedure, which may however not be compatible with the adiabatic manipulation of the potential.

\paragraph{\bf Non-adiabatic recombiner}

A way to avoid some limitations of the adiabatic Josephson beam splitter is to induce a fast interference between the condensates. This is done by decreasing the well spacing and the barrier height non-adiabatically with respect to the timescale of the motion. The clouds are abruptly accelerated towards the barrier and after an adjustable time $t_\mathrm{BS}$, the barrier is raised to separate the atoms for counting (see Fig.~\ref{Fig:BSCalib}d). Starting again with a symmetric superposition with $\phi_i=\pi/2$, the parameters of the recombiner trap and the time $t_\mathrm{BS}$ were optimized to yield the highest imbalance (see Fig.~\ref{Fig:BSCalib}e). The optimum was found in the same recombination double-well as for the first method, however for a longer time $t_\mathrm{BS}=2.25$\,ms. 

At any time $t_\mathrm{BS}$, the phase-dependent imbalance results from the interference between the parts of the wave-packet that are transmitted and reflected on either side of the semi-reflective potential barrier (see Ref.~\cite{Berrada2013} and SM). From a simple model where interactions are neglected and the double well is approximated by a 1D square potential barrier (see SM), it appears that the best contrast is achieved when 1) the kinetic energy of the wave packets is of the order of the barrier height $V_0$ and 2) the barrier is narrower than the typical tunneling depth $d_\mathrm{t} = \sqrt{2\hbar^2/(mV_0)}$, where $m$ is the mass of one atom. As can be seen from Fig.~\ref{Fig:SchematicsPotentials}b, the experimentally optimized double well precisely implements condition 1), while the spacing between the potential minima is approximately twice as large as the tunneling depth $d_\mathrm{t} = 0.5~\mathrm{\mu m}$ corresponding to the barrier height.     

The dynamics of the wave function in the double-well potential is fairly intricate but still can be captured by mean-field simulations. Note that conversely to the Josephson recombiner, the density profile exhibits a complex structure due to the multiple reflections and transmissions in the double-well potential. Simulations suggest that the optimal duration $t_\mathrm{BS}$ to achieve maximal imbalance corresponds to a turning point of the classical center-of-mass oscillations in the double well, for which the wave packets reach maximal separation (see SM).

Fitting the response $z(\phi_i)$ of the non-adiabatic recombiner at fixed $t_\mathrm{BS}=2.25$\,ms using the model of Eq.~\eqref{Eq:Anharmonic} restricted to the two lowest harmonics (see Fig.~\ref{Fig:BSCalib}), we find \mbox{$a_1=0.3\pm0.03$} and \mbox{$\eta=\left|a_2/a_1\right|=0.12\pm0.09$}, indicating a slightly weaker anharmonicity than for the Josephson recombiner. This behaviour is only partially captured by simplified 1D GPE simulations along the direction of splitting which predict $\eta=0.28$ (Josephson) and $\eta=0.23$ (non-adiabatic) respectively. One possible explanation for the fact that the non-adiabatic recombiner seems to be less affected by interactions is that the wavefunction is streched during the non-adiabatic motion, implying a lower mean-field energy.

Fig.~\ref{Fig:MZIfringes}c displays the Mach-Zehnder fringes obtained by varying the phase accumulation time. In contrast to the Josephson recombiner, the distribution of $z$ at all times for the non-adiabatic recombiner exhibits a clear double-peaked structure (see Fig.~\ref{Fig:MZIfringes}d) from which we extract $C = 42\pm3\%$ and $\sigma_\mathrm{RBS}=0.07$ using the model of Eq.~\eqref{Eq:BatmanDis}. This is less than expected from the time-dependent 3D GPE simulations (up to $\sim 70\%$, see Fig.~\ref{Fig:BSCalib}e). Interestingly, the simulated evolution of the output imbalance as a function of $t_\mathrm{BS}$ for the input phase $\phi_i=\pi/2$ (blue line in Fig.~\ref{Fig:BSCalib}e) displays roughly the same features as experimentally observed. This again suggests that an additional damping mechanism is at work. Fitting the experimental data with the result of the 3D GPE simulation with additional time shift and exponential damping prefactor yields $\Delta t=0.25$~ms and $\tau=5.4$~ms (red line in Fig.~\ref{Fig:BSCalib}e). This decay time is very close to the one we found for the Josephson recombiner (5.1~ms), suggesting that the damping mechanism is the same in both procedures.
%\textcolor[rgb]{1,1,0}{\textcolor[rgb]{1,0,0}{Note that although we worked in the same recombination double well for both recombiners, we simulated the BEC dynamics in slightly different recombination double wells to better match the data.}} In any case, the similar value of the extracted damping time may indicate that a relaxation mechanism similar to that observed with the Josephson recombiner is at work.

\section*{\bf Perspectives}

We have implemented two strategies for the phase-sensitive recombination of a BEC in a double-well potential. For the first time, a true analogue of an optical recombination beam-splitter is realized in a double-well based BEC Mach-Zehnder interferometer. The non-adiabatic recombiner, based on a fast modification of the double well yields a higher contrast than the Josephson recombiner. The latter, which relies on a manipulation of the potential which is adiabatic with respect to the motion of the atoms, ensures that the output state remains essentially in a superposition of the two lowest energy eigenstates of the potential. This is needed for example to further process the quantum state in a sequence of coherent operations. In contrast, the non-adiabatic recombiner produces a wave packet with a complex spatial structure of phase and density in each well. 

Currently, none of these recombination beam-splitter can allow for phase estimation better than the standard quantum limit \mbox{$\Delta \phi_\mathrm{SQL} = 1/\sqrt{N} = 0.03~\mathrm{rad}$}. We estimate the phase uncertainty from the Josephson recombiner $\delta \phi = \Delta z/ \left| \partial n / \partial \phi \right|_{\phi=0}\approx \sigma_z/C = 0.75~\mathrm{rad}$, while for the non-adiabatic recombiner we get $\delta \phi \approx 0.18~\mathrm{rad}$. Presently, the performance of both recombiners seems to be limited by an unknown relaxation mechanism which is responsible for the fast damping of the Josephson oscillations between the two wells. This damping, which is the subject of ongoing research in our group, is not captured by a 3D mean-field description. We conjecture that it is linked to the coupling between the coherent transverse dynamics and the longitudinal excitations of our elongated BECs. In our setup and within our experimental parameters, the atoms are in the 1D quasi condensate regime~\cite{Petrov2000}, where the atoms occupy only a few (typically: two) modes in the transverse direction of the double well, while they can access many longitudinal modes which form a "bath" the energy could decay to. One way to reach a better sensitivity to small phase shifts would be to achieve a higher contrast $C$. This may require the use of optimized trap manipulation protocols, in particular to operate on a timescale short with respect to these relaxation mechanisms~\cite{Grond2009,Grond2009b,Grond2010}.

These recombiners are necessary tools for the coherent manipulation of superpositions of external states, as needed for example to measure inertial forces. Since they allow mapping the relative phase between the two modes of the BEC into an atom number difference, these recombination protocols will greatly benefit from the precise atom counting methods already available~\cite{Esteve2008,Bucker2009,Lucke2011}. Furthermore, they allow taking advantage of interactions during the recombination step, which opens the way for quantum-enhanced interferometry with superpositions of external modes. While the sensitivity of the phase estimation based on the tof-recombination method is fundamentally bounded by $\delta \phi \propto N^{-2/3}$ regardless of the input state~\cite{Chwedenczuk2012}, recombiners based on atom counting are expected to be ultimately bounded by the Heisenberg scaling $\delta \phi \propto 1/N$~\cite{Pezze2005,Grond2011} only. Eventually, our recombiners complement the atom-optics toolbox for the manipulation of superpositions of motional states of a BEC and can be used for the tomography of the many-body wavefunction.  

%
%
%Finally, let us stress that the use of the tools 
%devised in this work in real life applications is clearly foreseeable, 
%e.g. a compact and ready to be used in quantum technologies BEC 
%has been recently presented in Ref.~\cite{C.2013}.

\subsection{Acknowledgments}
\begin{acknowledgments}
We are grateful to Thomas Betz, Stephanie Manz and Aurélien Perrin for earlier work on the experiment. T.B. and R.B. acknowledge the support of the Vienna Doctoral Program on Complex Quantum Systems (CoQuS).
J.-F.S. acknowledges the support of the Austrian Science Fund (FWF) through his Lise Meitner fellowship (M 1454-N27). This research was supported by the European STREP project QIBEC (284584), the European Integrated project AQUTE (247687) and the FWF projects SFB FoQuS (SFB F40) and CAP (I607-N16). B. J-D is supported by the Ramon y Cajal program. The authors acknowledge financial support by grants FIS2014-54672-P Generalitat de Catalunya and FIS2011-24154 from MICINN (Spain).
\end{acknowledgments}

%\bibliographystyle{apsrev4-1}
%\bibliography{RecombinersRef}

%

\clearpage

\appendix

\section{Supplemental Material: further details on the recombiners}

In this supplemental material, we provide more details on both recombining procedures 
explored in the letter. 
 
\subsection{Josephson recombiner}

The mean-field description of the bosonic Josephson junction 
can be written in terms of the population imbalance $z(t)=(N_L(t)-N_R(t))/N$ and 
its canonical conjugate, the phase difference between the two wells, 
$\phi(t)=\phi_R(t)-\phi_L(t)$. As derived in Ref.~\cite{Raghavan1999} the classical 
Hamiltonian governing their dynamics reads, 
\begin{equation}
H = \Lambda z^2 -\sqrt{1-z^2}\cos \phi
\end{equation}
with $\Lambda  = NU/(2J)$ and 
\begin{eqnarray}
J&=&- \int \left( {\hbar^2 \over 2m} \nabla \phi_L \nabla \phi_R + \phi_L \phi_R V\right) d \vec{r} \nonumber \\
U&=& g \int \phi_L^4 d\vec{r}
\end{eqnarray}
with $g={4 \pi \hbar^2 a_s \over m}$ and $a_s$ the s-wave scattering length. 
$\phi_{L(R)}$ are modes localized on the left (right) well and $V$ is the double-well potential. 

Here we considered the test case of an initial $\pi/2$ phase 
for the proof of principle of the recombiners. Repeating the whole procedure 
for different values of the initial phase, we observed a sine-like 
dependence of the final imbalance, as displayed in Fig.~2c. 
The signal exhibits a characteristic anharmonicity, which is also 
captured by mean-field simulations. It indicates that the slope 
of $z(\phi)$ is steeper when $\phi$ is close to $\pi$ than to when 
$\phi$ is close to zero, and can be attributed to atomic interactions. 
It is linked to the fact that classically, in presence of interactions, 
the initial condition $(z=0,\phi=\pi)$ is a hyperbolic fixed point 
(see Fig.~3b).

\subsection{Non adiabatic recombiner}

%\begin{figure}[t] %\hspace*{-10px}
%\centering
%\includegraphics[width=\linewidth]{./figures/NonAdiabaticFringe.png}
%\caption{\textbf{Mach-Zehnder interferometric fringes obtained with the 
%non-adiabatic recombiner.} Normalized population imbalance $z$ as a function 
%of the phase accumulation time $t_\phi$ obtained using the non-adiabatic 
%recombiner (note that the scale is the same as in Fig.~\ref{Fig:JBSfringes} 
%for both axes). The contrast of the fringes is higher and the anharmonicty 
%is less obvious than with the Josephson recombiner. Grey dots: imbalance 
%of individual experimental realizations. Black dots: ensemble average 
%$\left\langle z\right\rangle$ at each phase accumulation time (the error 
%bars denote $\pm$ one standard error of the mean). Red: fit with a 
%theoretical model for phase diffusion. 
%\label{Fig:FBSfringes}} 
%\end{figure}	

\subsubsection{Principle}

Similarly to the Josephson recombiner we have repeated the recombining 
sequence for different values of the initial phase, we again observed a 
sine-like dependence of the final imbalance, as displayed in Fig.~2f.

The phase-dependent dynamics arises from the interference between the 
transmitted and the reflected wavepackets on either side of the barrier, 
like in a half-silvered mirror. Neglecting interactions, the symmetry 
of the potential and the linearity of the Schrödinger equation ensure that 
the population imbalance 
$z(t) \equiv \int_{-\infty}^{0} \left|\psi\right|^2\mathrm{d}x 
- \int_{0}^{\infty} \left|\psi\right|^2\mathrm{d}x$ obeys at each time
\begin{align}
z(t) &= C(t) \sin \phi_i, \\
\text{where}~C(t) &= 2\int_{0}^{\infty} \mathrm{Im} 
\left[ \psi_R^*(x,t) \psi_L(x,t)  \right] \mathrm{d}x, \label{Eq:ContrastBS}
\end{align}
the time-dependent contrast $0 \leq C(t) \leq 1$ is independent of $\phi_i$ 
and is determined by the mode-matching between the transmitted and the 
reflected wave packet on either side of the barrier. A simple model assuming 
a square barrier predicts that the best contrast is achieved when 1) the 
height $V_0$ of the potential barrier is of the order of the kinetic 
energy of the atoms and 2) the barrier is narrow enough ($d \leq \hbar/\sqrt{2mV_0}$)) 
for the atoms to tunnel through the barrier. This condition is approximately 
realized in the double-well potential for which the highest contrast 
is achieved (see Fig.~1b). 

\subsubsection{Square-barrier model}

The simplest model of the non-adiabatic recombiner consists in neglecting interactions and approximating the double-well potential with a one-dimensional square barrier. A wave packet impinging on the barrier is split between a transmitted and a reflected wave. For a plane matter wave of momentum $\hbar k$ (energy: $\hbar^2 k^2/(2m)$) impinging on a potential barrier of height $V_0$ and size $d$, the transmission coefficient reads
	\begin{align}
	T & = \frac{4 \epsilon \left(\epsilon-1\right)}{4 \epsilon \left(\epsilon-1\right)+ \mathrm{sin}^2 \left[\sqrt{\left(\epsilon-1\right)}d/L\right] } \quad	\text{if $E>V_0$}\\
	T  & = \frac{4 \epsilon \left(1-\epsilon\right)}{4 \epsilon \left(\epsilon-1\right)+ \mathrm{sinh}^2 \left[\sqrt{\left(1-\epsilon\right)}d/L\right] } \quad	\text{if $E<V_0$}
	\end{align}
where $\epsilon=V_0 /(\hbar^2 k^2/2m)$ is the kinetic energy of the plane wave in units of the barrier height and $L(V_0)=\hbar/\sqrt{2 m V_0}$ is the tunneling length associated to the energy $V_0$. $L$ corresponds to the extension of a wave packet of kinetic energy $V_0$, and is a typical measure of the penetration depth of an evanescent matter wave into a potential barrier of height $V_0$ at low energy~[68].

\begin{figure}[h] %\hspace*{-10px}
\centering
\includegraphics[width=\linewidth]{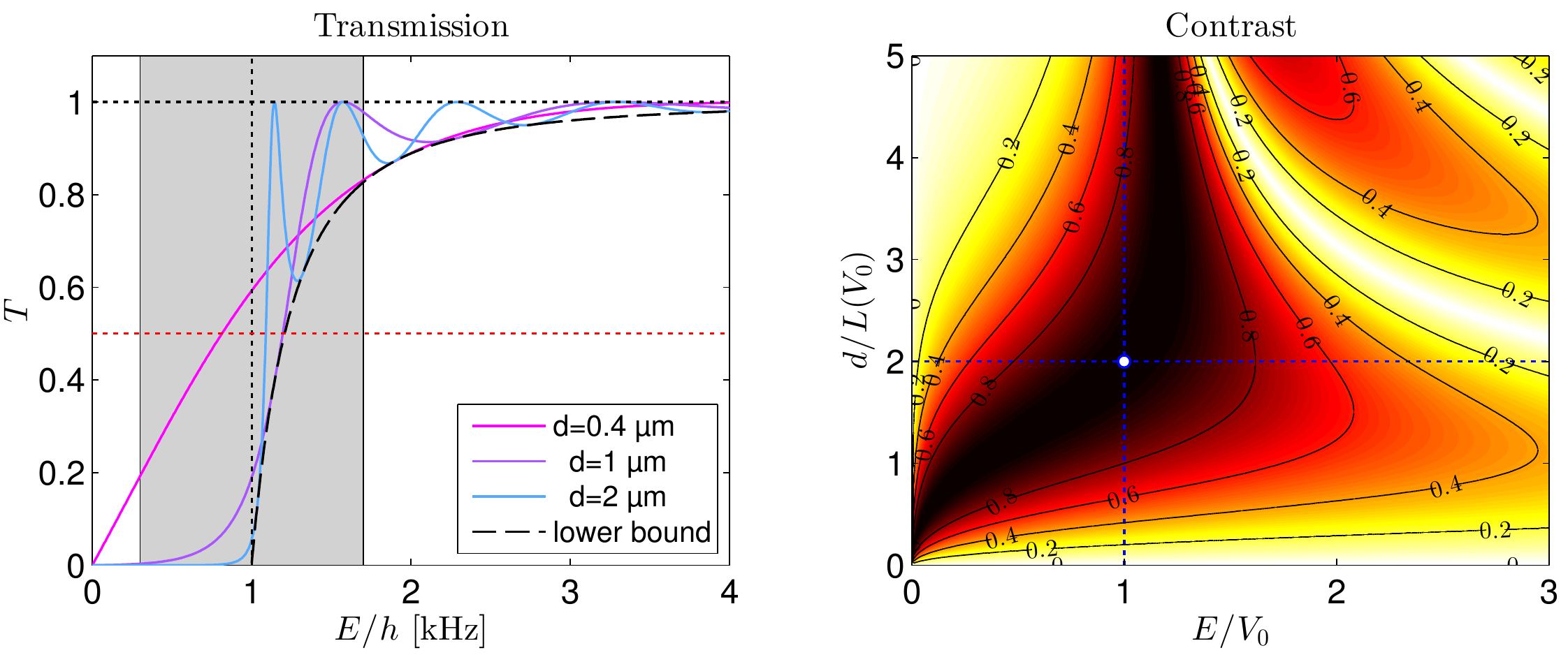}%
\caption{ \textbf{Transmission and contrast of the square beam splitter.} \textbf{Left:} Transmission probability for a plane wave of energy $E$ impinging on a potential barrier of height $V_0/h=1$~kHz and width $d=0.4,~1$ or 2~$\mathrm{\mu m}$.  Note the oscillations of $T$ associated with the transmission resonances for $E>V_0$. Dashed black line: lower bound for $T$ in the classical regime ($E>V_0$). Gray shaded area: uncertainty on the kinetic energy of the initial state in the double well $\Delta E = \pm \hbar \omega/2$. The red dotted line corresponds to the transmission of a semi-reflective mirror $T=0.5$. \textbf{Right:} Contrast of a square beam splitter when two plane matter waves of opposite momentum and equal intensities are impinging on it. High contrast can be achieved in the tunneling regime ($E<V_0$) provided the barrier is sufficiently narrow and in the classical regime when $E \approx V_0$. At higher energy, secondary maxima can be observed when a transmission resonance is reached. 100\% contrast is obtained for $E=V_0$ and $d=2L$ (white point). \label{Fig:SquareBarrier}}
\end{figure}

Figure~\ref{Fig:SquareBarrier}, left panel, shows how the transmission probability $T$ depends on the energy and the barrier. Two regimes must be distinguished:

\begin{itemize}
	\item $E>V_0$ corresponds to a situation where classically, the particles would pass over the barrier. Quantum mechanically, the wave packet is partly transmitted and partly reflected. The transmission probability oscillates between $\left[1+4\epsilon\left(\epsilon-1\right)\right]^{-1}$ (thick black dashed line) and 1. Transmission resonances occur whenever the energy of the incoming wave corresponds to the existence of a standing wave in the barrier. To build a 50:50 beam-splitter, one must achieve $T=0.5$. The lower bound for $T$ imposes the (necessary) condition
	\begin{align}
	E \leq \frac{1+\sqrt{2}}{2}V_0.	
	\end{align}
	In other words: a 50:50 beam-splitter can only be achieved in the classical regime ($E>V_0$) if the energy is of the order of the barrier height ($V_0 \leq E \lesssim 1.2 \times V_0 $).
	
	\item $E<V_0$ corresponds to a situation where the atoms can only tunnel through the barrier. The transmission probability is a monotonically decreasing function of $E$. $T=0.5$ can only be achieved if
	\begin{align}
		d < 2L(V_0).
	\end{align}
	This second condition simply means that in the tunneling regime, the transmission drops when the barrier is larger than the penetration depth.
\end{itemize}

Still, $\left\langle E\right\rangle \approx V_0$ is not sufficient to achieve a high contrast. The mode-matching condition of Eq.~\eqref{Eq:ContrastBS} shows that in order for $C$ to be large, there must be a good overlap between the reflected and the transmitted wave on each side of the barrier. 

In the case of a square barrier, we can derive an explicit expression for the contrast from the model of Ref.~\cite{Cohen-Tannoudji1977b}. Assuming that two plane waves of equal intensity and opposite momentum are impinging on the square barrier, the contrast reads
\begin{align}
	C & = \frac{4\sqrt{\epsilon \left(\epsilon - 1\right)} \sin \left[\sqrt{\left(\epsilon-1\right)}d/L\right] }{4 \epsilon \left(\epsilon - 1\right) + \sin^2 \left[\sqrt{\left(\epsilon-1\right)}d/L\right]} \quad	\text{if $E>V_0$}, \nonumber \\ 
	C & = \frac{4\sqrt{\epsilon \left(1-\epsilon \right)} ~ \mathrm{sinh} \left[\sqrt{\left(1-\epsilon\right)}d/L\right] }{4 \epsilon \left(1-\epsilon \right) + ~\mathrm{sinh}^2 \left[\sqrt{\left(1-\epsilon\right)}d/L\right]} \quad	\text{if $E<V_0$}. \label{Eq:ContrastSquareBS}
\end{align}
The result is displayed in Fig.~\ref{Fig:SquareBarrier}, right panel. As expected, the maximal contrast $C=1$ is achieved when
\begin{align}
	E &= V_0, \\
	\mathrm{and} \quad d &= 2L(V_0) = \sqrt{\frac{2\hbar^2}{mV_0}}.
\end{align}

When the kinetic energy is larger than the barrier height, the contrast is approximately equal to $V_0/E$.  In the tunneling regime ($E<V_0$), high contrast can be achieved, provided the barrier is made narrow enough. For a given energy $E$ of the incoming waves, taking the limit $d \to 0$ imposes that $V_0$ must diverge like $1/d$ to ensure a contrast of unity. This corresponds to the limit of an ideal $\delta$ potential, or, in optics, to an infinitely thin half-silvered mirror.

In practice however, the wave packets are not plane waves, they have instead a finite momentum spread which is non-negligible compared to $V_0$ (gray shaded area in Fig.~\ref{Fig:SquareBarrier}). It means that the atoms tunnel through the barrier as much as they cross it classically. The potential barrier in our double wells also has a finite extension, which is always comparable to the width of wave packets (see Fig.~1b). 

%Furthermore, in our rf dressed double wells, the well spacing and the barrier height are not independent\footnote{At least, they cannot be tuned independently with only one control parameter (the dressing amplitude). It has not yet been checked whether adding a new degree of freedom, such as the RF dressing detuning, could allow tuning $d$ and $V_0$ independently.}: $V_0$ increases roughly like $d^4$. From simulations of the double wells, we find that it should be possible to fulfill $d<2L$ up to $\mathrm{RF}_\mathrm{Amp}=0.6$. Interestingly, the best contrast was precisely achieved in the double well $\mathrm{RF_{Amp}}=0.55$, for which simulations predict that the barrier height is equal to the potential energy of the clouds (see Fig.~\ref{Fig:SymPot}). 

\subsubsection{Dynamics of the BEC in the beam-splitter}

\begin{figure}[t]
\centering
\includegraphics[width=\linewidth]{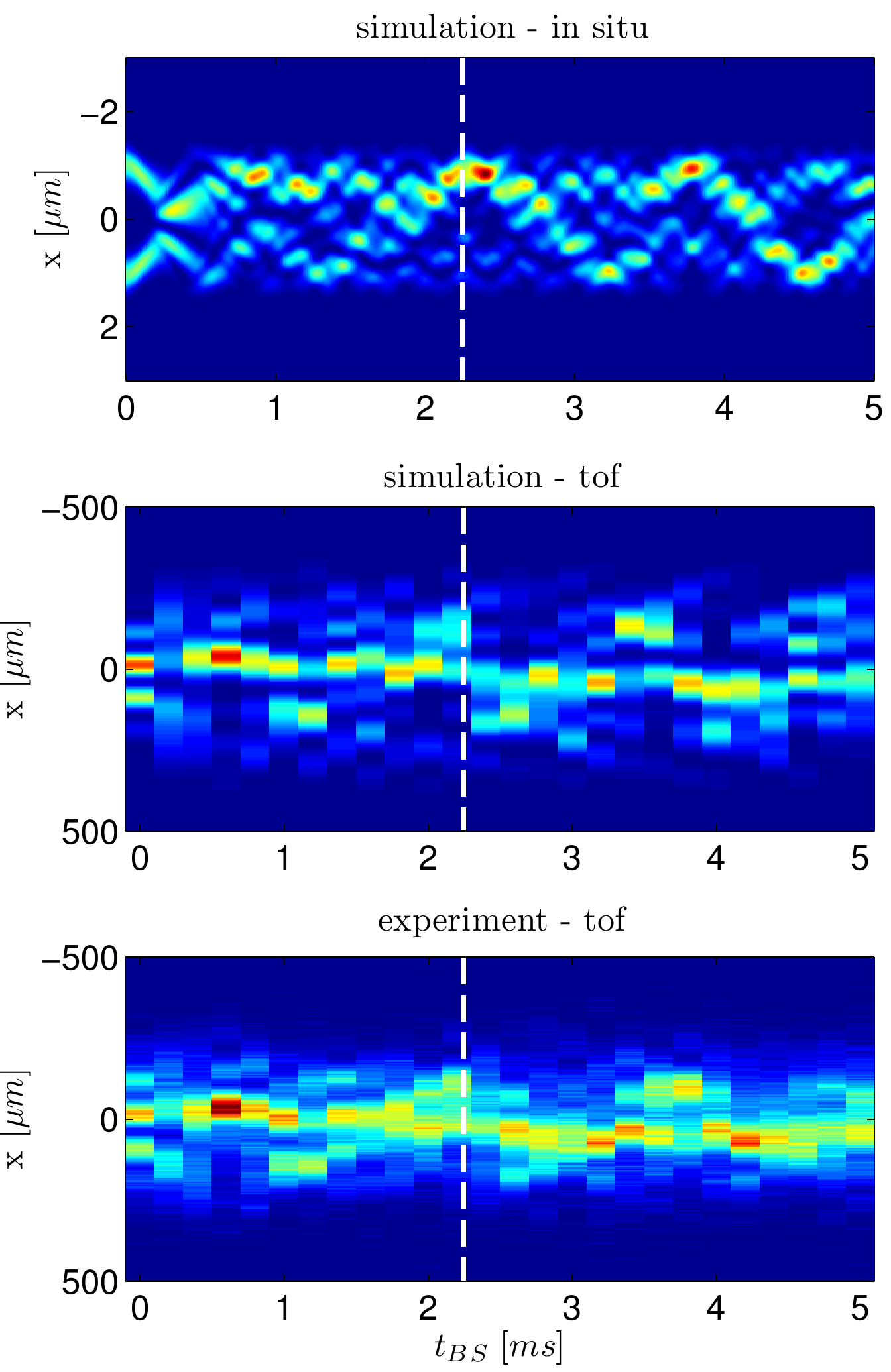}
\caption{\textbf{Dynamics of the condensates in the recombiner.} Using the 
phase-shifter, a state with a phase of $2.13$\,rad is prepared. Its dynamics 
in the recombiner is monitored by time-of-flight imaging. From top to bottom: 
simulation of the \textit{in situ} density profile;  simulation 
of the momentum distribution scaled to the 46ms time of flight, including 
the finite imaging resolution; measured density profiles after expansion. 
The white dashed line corresponds to $t_{BS} = 2.25$\,ms giving maximum contrast. 
We checked the agreement between data and simulation for 8 initial values of the phase.}
\label{Fig:BSsim}
\end{figure}

To confirm our understanding of the dynamics of the wavefunction, we 
prepared states with 8 different relative phases and abruptly reduced 
the splitting similarly to the non-adiabatic recombination. After holding 
the atoms for an adjustable time $t_\text{BS}$, we eventually switched 
off the potential instead of separating the clouds. Since for our 
parameters, the transverse density distribution after time of flight is 
almost homothetic to the initial transverse momentum distribution, this 
gives us access to the momentum distribution at any time in the 
recombiner. We compared it to the results of the simulation and 
found qualitative agreement for all phases (see Fig.~\ref{Fig:BSsim}).

\begin{figure}[h] %\hspace*{-10px}
\centering
\includegraphics[width=\linewidth]{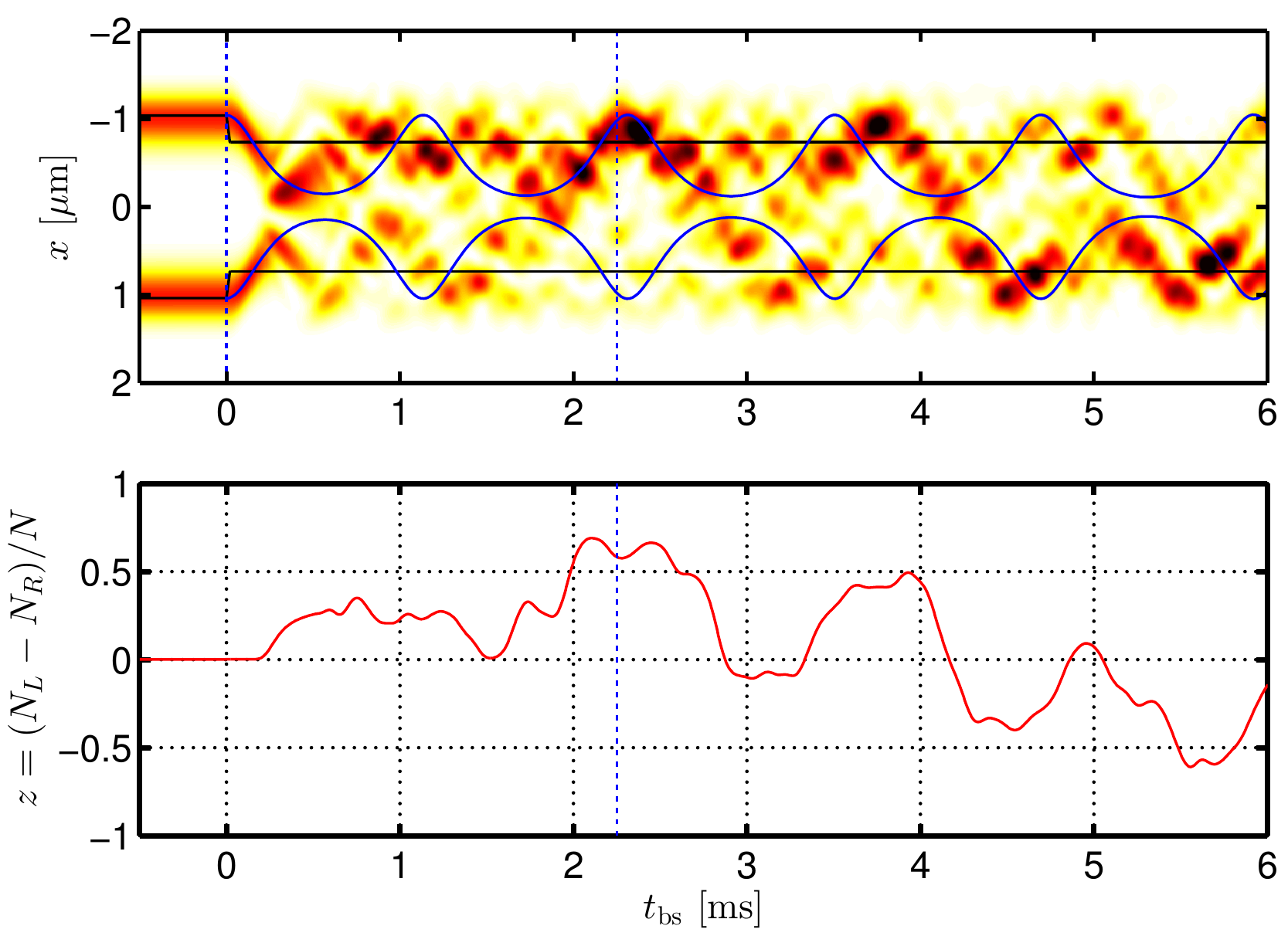}%
\caption{ \textbf{Dynamics of the BEC in the recombiner.} Density profile of the BEC in the recombiner (\textbf{top}, 1D GPE simulation. $\phi_i=\pi/2$) and corresponding population imbalance (\textbf{bottom}). At $t=0$, the splitting is abruptly reduced. An oscillating feature emerges in the complex dynamics of the density pattern. It roughly corresponds to the center-of-mass oscillations of the atoms in each well (continuous blue lines). At regular intervals, imbalance builds up. Vertical dashed line: time at which the barrier is raised in the normal recombining procedure in order to separate the atoms for counting ($t_\mathrm{BS}=2.25$~ms). It corresponds to the turning point of the second oscillation period.\label{Fig:ClassicalDynamics}}
\end{figure}

Solving the classical equations of motion for a point-like particle initially at the same position as the center of mass of one of the BEC (see Fig.~\ref{Fig:ClassicalDynamics}), we see anharmonic oscillations in each single well at the period $T_\mathrm{anh} \approx 1.2$~ms ($\nu_\mathrm{anh} \approx 800$~Hz, note that the oscillating pattern in the GPE simulations has a higher frequency than the classical oscillation).

For an initial phase of $\pi/2$, imbalance seems to build up after an integer number of center-of-mass oscillations. The final imbalance can be maximized by separating the clouds when they are at a classical turning point, i.e.\ when the distance between the clouds is maximal. Indeed, the evolution of the final imbalance shows a sequence of bumps and dips spaced by roughly 1 ms. How many center-of-mass oscillations are necessary to reach the maximum imbalance is not obvious from the GPE simulations, but in practice, the pattern we observed experimentally (see Fig.~2e) suggests that already after a few ms, unknown relaxation mechanisms completely damp the coherent evolution.

\end{document}